# On transversely isotropic mechanical properties of vacancy defected carbon nanotubes

## S. I. Kundalwal and Vijay Choyal

*Applied and Theoretical Mechanics Laboratory, Discipline of Mechanical Engineering,*

*Indian Institute of Technology Indore, Simrol, Indore 453 552, India*

## ABSTRACT

Molecular dynamics (MD) simulations with Adaptive Intermolecular Reactive Empirical Bond Order (AIREBO) force fields were conducted to determine the transversely isotropic elastic properties of carbon nanotubes (CNTs) containing vacancies. This is achieved by imposing axial extension, twist, in-plane biaxial tension and in-plane shear to the defective CNTs. The effects of vacancy concentrations, their position and the diameter of armchair CNTs were taken into consideration. Current results reveal that vacancy defects affect (i) the axial Young's and shear moduli of smaller diameter CNTs than the larger ones and decrease by 8% and 16% for 1% and 2% vacancy concentrations, respectively; (ii) the plane strain bulk and the in-plane shear moduli of the larger diameter CNTs more profoundly, reduced by 33% and 45% for 1% and 2% vacancy concentrations, respectively; and (iii) the plane strain bulk and in-plane shear moduli among all the elastic coefficients. It is also revealed that the position of vacancies along the length of CNTs is the main influencing factor which governs the change in the properties of CNTs, especially for vacancy concentration of 1%. The current fundamental study highlights the important role played by vacancy defected CNTs in determining their mechanical behaviors as reinforcements in multifunctional nanocomposites.

**Keywords**: Carbon nanotube, Vacancy defects, Molecular dynamics, Transversely isotropic, elastic properties

E-mail address: kundalwal@iiti.ac.in (S.I. Kundalwal).

# 1 Introduction

In view of their extraordinary mechanical and physical properties, carbon nanotubes (CNTs) have emerged as one of the most promising nano-reinforcements that can also be used to tailor the properties of polymers [1–7]. With the rapid development in nanocomposites, it becomes essential to characterize the mechanical and thermal properties of CNTs. Due to the inherent limitations of fabrication and purification processes, CNTs typically contain different defects such as atom vacancies, doping, substitutional impurities, Thrower-Stone-Wales (TSW), and hybridization. In addition, intra- and inter-molecular junctions are other special defects which include one or more of the basic defects classified above. Under certain circumstances, defects in CNTs are deliberately introduced by irradiation with energetics particles or by chemical processing for the purpose of tailoring their properties to suit the requirements of the application. When these defects evolve, the local change in their chemical bond order and conformation alters the mechanical and thermal properties of CNTs as well as their nanocomposites.

A number of studies have been conducted to investigate the influence of defects on the elastic properties of CNTs. An earlier attempt was made by Ajayan et al. [8] to study the unstable single vacancy containing three hanging bonds. Belytschko et al. [9] studied the influence of missing atoms and TSW defects on the fracture behavior of CNTs. Their study indicates that Young's modulus of a CNT has very little dependence on the presence of TSW defects. A quantum mechanics study by Troya et al. [10] revealed that defects can result in the decrease of ultimate bearing capacity of CNTs. Lu and Pan [11] carried out a systematic calculation of single vacancies in chiral CNTs. They further obtained the relationship between the vacancy formation energies and the radii of different CNTs. Hao et al. [12] investigated the influence of defects on the elastic properties of (7, 7) and (12, 12) CNTs through MD simulations and reported respective reductions of 4% and 5% in their Young's moduli by adding one vacancy defect to their structures. Yuan and Liew [13] carried out a MD study to investigate the vacancy defect reconstruction and elastic properties of armchair (5, 5) and (10, 10) CNTs with different defect ratios. When the vacancy defect achieves a certain ratio, they observed that a sudden slowdown in Young's moduli for a given vacancy defect ratio. Based on the continuum mechanics method, Chen et al. [14] developed three-dimensional finite element (FE) models to study the effects of different atom vacancy defects on Young's and shear moduli of armchair and zigzag CNTs. Their parametric study indicates that the CNT diameter significantly affects the elastic constants of CNTs among all other factors. Fefey et al. [15] investigated the effect of position of defects on Young's modulus of a CNT using MD simulations. Their study indicated that for an enclosed defect with the same shape in a CNT structure, its position does not cause any change in Young's modulus. However, as the number of defects increases, the predicted Young's modulus was found to decrease. FE models of three types of defected CNTs (doping with Si atoms, carbon vacancy and perturbation of the ideal location of the carbon atom) were developed by Ghavamian et al. [16]. Their results showed that the existence of any kind of imperfection in the CNTs leads to lower stiffness values. Through MD simulations of CNTs either with a single TSW defect or vacancy defect, Talukdar and Mitra [17] reported the degradation of all elastic properties of CNTs. Likewise, Sharma et al. [18] concluded that for the same defect densities, vacancy defect results in the deterioration of the tensile strength of CNTs much more since such type of defects create holes or voids in the CNTs at which their failure can initiate. Zhou and Liao [19] observed that the patterns of atomistic crack propagation and fatigue behavior of CNTs are influenced by the type and orientation of TSW defects, inter-defect distance and the magnitude of externally applied stress. Rafiee and Pourazizi [20] investigated the influence of vacancy and TSW defects on the elastic properties of CNTs using nanoscale continuum mechanics based FE model. They determined Young's moduli of defected CNTs and reported that the importance of vacancy defect is considerably higher than that of TSW defects implying on the drawback of chemical functionalization process. Sakharova et al. [21] performed the FE simulations to study the mechanical behaviour of non-chiral and chiral CNTs containing different percentage and types of vacancy defects under tensile, bending and torsional loadings. Their results revealed that the three rigidities decrease with the increase of the percentage of vacancies, and Young's and shear moduli of CNTs are sensitive to the presence of vacancies. Alian et al. [22] reported that the mechanical performance of polycrystalline CNTs is largely governed by the grain boundary defects but not the size of grain boundaries using MD simulations. Bocko and Lengvarský [23] studied the buckling behaviour of defected CNTs using FE method and reported that the critical buckling force of CNT decreases as the number of defects in it increases. MD simulations were carried out by Shahini et al. [24] to assess the tensile properties of three coiled (3, 3), (4, 4), and (5, 5) CNTs. The results indicated that TSW defects are necessary for thermodynamic stability of the coiled CNTs.

When CNTs are used as reinforcements in nanocomposite systems, the above-mentioned characteristics of defects may lead to conflicting contributions to their overall performance. For example, Joshi et al. [25] applied micromechanics simulation of CNTs containing pinhole defects to analyze the degradation of elastic properties of CNT-based nanocomposite. They observed that the presence of pinhole defects reduces the elastic properties of CNTs and their nanocomposites. Using MD simulations, Yang et al. [26] investigated the effect of TSW defects on the elastic stiffness of polymer nanocomposites. They reported that the TSW defects degrade the elastic properties of CNTs and improve the interfacial adhesion between the defective CNTs and polymer molecules. In another effort, Sharma et al. [27] observed that the elastic properties of armchair CNTs and their nanocomposites degrade with increasing CNT diameter and the number of defects in the CNTs. Mahboob and Islam [28] examined the effect of defective CNTs on the interfacial and elastic properties of CNT/polyethylene composite via MD simulations and showed that the vacancies in CNTs decrease the longitudinal Young's and shear moduli of the nanocomposite. Lv et al. [29] reported that the TSW and single adatom defected CNTs affect the elastic properties of their polypropylene nanocomposites. Likewise, Kumar et al. [30] reported that the vacancy and TSW defected CNTs reduce the elastic properties and failure strain of their epoxy nanocomposites.

Using CNTs as reinforcements to design and fabricate multifunctional composites [31–48] with desirable elastic properties necessitate a thorough understanding of the mechanical behavior of defective CNTs. It is well known that the mechanical behavior of CNTs is transversely isotropic when subjected to different loading conditions [2–5, 7]. The review of literature presented on defective CNTs clearly indicates that they are considered as isotropic while determining their elastic properties. To the best of current authors' knowledge, there has been no comprehensive study for determining the transversely isotropic elastic properties of vacancy defected CNTs. Therefore it is the objective of this study to determine the transversely isotropic elastic properties of vacancy defected CNTs. This is achieved through the quantitative evaluation of all five independent elastic moduli of perfect and defective CNTs with the aid of MD simulations. Selection of CNTs with vacancy defects herein is by the fact that they are the most common and significant among the typical defects experienced in CNTs [9, 13, 17, 18, 20, 49–51]. Such vacancies appear due to the chemical functionalization and electron irradiation of the CNTs. Functionalization can enhance the load transfer from the surrounding polymer to the reinforced CNTs. Functionalized CNTs also display unique properties that enable a variety of medicinal applications, including the diagnosis and treatment of cancer, drug design and discovery, biosensing, infectious diseases and central nervous system disorders, and applications in tissue engineering [52, 53]. Therefore, it becomes necessary to investigate the elastic properties of CNTs containing vacancies.

## 2 MD modeling of CNTs

MD is the most widely used modeling technique for the simulation of nanostructured materials because it allows accurate predictions of interactions between atoms and molecules at the atomic scale level. It involves the determination of the time evolution of a set of interacting atoms, followed by integration of the corresponding equations of motion [50]. The obvious advantage of MD over classical models is that it provides a route to dynamical properties of the system: transport coefficients, time-dependent responses to perturbations, rheological properties and spectra, thermo-mechanical properties and many more unique characteristics. Therefore, MD simulations were conducted in the current study to determine the elastic properties of CNTs at the atomic scale. All MD simulations runs were conducted with large-scale atomic/molecular massively parallel simulator (LAMMPS) [54] and the molecular interactions in CNTs are described in terms of Adaptive Intermolecular Reactive Empirical Bond Order (AIREBO) force fields [55].

Determination of transversely isotropic elastic properties of CNTs was accomplished via the use of the strain energy density-elastic constant relations. The equivalent continuum structure of a CNT is assumed to be annular cylinder by considering its effective wall thickness as 3.4 Å [12, 16, 17, 20, 26, 27, 49, 51, 56], and its cross-sectional area and volume are determined as $A = 2\pi R_i t$, $V = 2\pi R_i t L$ where $R_i$ is the inner radius, L is the length of a CNT and t is it's wall thickness. Note that the strain energy variations in a CNT atomistic structure and its equivalent solid continuum due to deformation are assumed to be same, and accordingly, continuum-based relations were used to determine the elastic coefficients. The methodology involves applying a load to a CNT and then computing the energy due to interatomic interactions as a function of the imposed load. A direct transformation to continuum properties is then made by assuming that the potential energy density of discrete atomic interactions is equal to the strain energy density of the continuous substance occupying a CNT volume. This approach has been widely adopted by several researchers [2, 3, 16, 18, 20, 22, 27] and they validated their results with those predicated by different approaches.

First, the initial structures of CNTs were prepared. Then, the initial structures of the CNTs were optimized to their minimum energy configurations using the conjugate gradient algorithm. The minimized structure of a CNT was considered to be optimized once the change in the total potential energy of the system between subsequent steps is less than $1.0 \times 10^{-10}$ kcal/mol. Subsequently, MD simulations were performed in the constant temperature and volume canonical (NVT) ensemble at a time step of 1 fs for 50 ps to equilibrate the CNT structures. The velocity Verlet algorithm was used to integrate the equations of motion in all MD simulations.

The vacancies in the CNTs were introduced by removing carbon atoms from the pristine CNTs. Note that taking advantage of the possibility to create a controlled vacancy concentration in CNTs by electron or ion irradiation process [57], one can alter their mechanical properties. The vacancy concentration ($\rho$) is defined to express the amount of incorporated defects as follows:

$$\rho = \frac{\text{number of removed atoms}}{\text{total atoms in a pristine CNT}} \times 100 \qquad (1)$$

To create a vacancy defect, carbon atoms were removed from the middle portion of CNTs as shown in Fig. 1. This figure shows the structures of pristine and defective armchair (5, 5) CNTs. Figures 1(a) and 1(b) show the structures of defected (5, 5) CNTs with 2 and 4 vacancy states corresponding to 1% and 2% vacancy concentration, respectively. MD simulations were performed on five types of armchair CNTs: (5, 5), (10, 10), (15, 15), (20, 20) and, (25, 25). The corresponding diameters of these CNTs are 6.781 Å, 13.563 Å, 20.344 Å, 27.125 Å and 33.339 Å, and their other geometrical parameters are listed in Table 1. The five types of zigzag CNTs [(0, 9), (0, 17), (0, 26), (0, 35), and (0, 43)], nearest in diameters to the corresponding armchair CNTs, were also considered. The corresponding diameters of zigzag CNTs are 7.047 Å, 13.337 Å, 20.358 Å, 27.405 Å and 33.669 Å. Unless otherwise mentioned, the vacancies are considered to exist in the middle portion of CNTs. The findings and inferences from existing MD simulations are also given.

## 2.1 Elastic moduli of CNTs

To determine the five independent elastic moduli, the following four loading conditions were imposed: axial tension for axial Young's modulus ($E_1$) and major Poisson's ratio ($\nu_{12}$), torsional moment for axial-shear modulus ($G_{12}$), in-plane biaxial tension for plane strain bulk modulus ($K_{23}$), and in-plane shear for in-plane shear modulus ($G_{23}$). Schematics of these loading conditions are demonstrated in Fig. 2. The written equations underneath the figures indicate the respective strain energy densities (U) stored in the CNTs due to the applied load. Defined strain increments of 0.1% were applied to the CNTs followed by the potential energy minimization process. We deformed a CNT in small strain increments (0.1%) to equilibrate its deformed state over an interval of 10 ps to reduce the effect of fluctuations. The 10 ps is quite adequate to equilibrate the deformed state of a CNT. Note that the fluctuations in the temperature and potential energy profiles are less than 1% when the system reaches equilibrium after about 5 ps [58, 59]. The loading steps were repeated until the total strain reached up to the failure strain. Under such deformations, the change in potential energy is equivalent to the stored strain energy in the CNTs and the stress is calculated from energy-strain curve as

$$\sigma = \frac{1}{V}\frac{dE}{d\varepsilon} \tag{2}$$

where σ is the longitudinal stress, V the volume of a CNT, ε is the strain and E is the stored strain energy in the nanotube [17]. The elastic constant of each case of deformation was obtained from the deformation energy density–elastic constants relations and from the initial slope of the stress-strain curve.

In case of axial tension, the atoms at the bottom end of a CNT were fully constrained, while at the top end an incremental axial displacement was applied. At the top end of a CNT, the atoms were also constrained in the transverse plane to prevent its bending at high loads. Since the local deformations of the carbon atoms are similar due to the symmetry of its atomic structure during the axial loading, an equivalent structure of a CNT with length L can be used for the energy analysis [2]. Accordingly, the strain energy density of a CNT can be expressed as

$$U = \frac{1}{2}E_1\bar{\varepsilon}^2 \tag{3}$$

in which U is the strain energy per unit volume and is given by,

$$U = \frac{\Delta E}{AL} \tag{4}$$

where ΔE is the increment of the potential energy, A is the cross-sectional area of a CNT, $\bar{\varepsilon} = \Delta L/L$ is the axial strain of a CNT, and ΔL is the increment of a CNT length.

It may be noted that the CNTs are assumed as thin cylindrical shells, except CNT (5, 5) which was considered as solid cylinder [2]. Substituting the proper values in Eq. (3), the axial Young's modulus ($E_1$) can thus be determined.

The concept of modeling a CNT molecular structure as an effective "stick-spiral" system was used to determine the Poisson's ratio ($\nu_{12}$) and the detailed methodology can be found in Refs. [60, 61]. Using the following definition, the Poisson's ratio ($\nu_{12}$) of a CNT can be obtained,

$$\nu_{12} = -\frac{\bar{\varepsilon}'}{\bar{\varepsilon}} \tag{5}$$

where $\bar{\varepsilon}'$ and $\bar{\varepsilon}$ are the respective circumferential and axial strains.

In case of twisting moment, the atoms at the bottom end of a CNT were fully constrained while at the top end an incremental tangential displacement was applied. The mechanical test for the local deformation of a CNT subjected to a torque is more complex than that associated with axial tension. In this case, all atoms were constrained radially in the transverse plane to maintain the presumed cylindrical surface of a CNT. The strain energy density of a CNT can be expressed as

$$U = \frac{G_{12}\phi^2 J}{2L} \tag{6}$$

in which $\phi$ is the angle of twist of a CNT and J is the polar moment of inertia; viz,

$$J = \frac{\pi}{32}[D^4 - (D - 2t)^4] \tag{7}$$

where D is the outer diameter of a CNT.

In case of in-plane biaxial tension, all atoms of a CNT were subjected to two-dimensional plane strain condition. In addition, two ends of a CNT were constrained axially so that its length remains unchanged, which satisfies the plane strain condition. Using the following definition [2, 5], the plane strain bulk modulus ($K_{23}$) of a CNT can be obtained,

$$U = 2K_{23}\bar{\varepsilon}'^2 \tag{8}$$

In case of in-plane shear condition, a CNT was subjected to in-plane shear at small strains in such way that its circular cross-section deform into an elliptical form in a similar fashion to the situation of a thin cylindrical shell, which is described as [2]

$$r = R(1 + \overline{\epsilon})\cos\theta\mathbf{i} + R(1 - \overline{\epsilon})\cos\theta\mathbf{j} \qquad \theta \in [0, 2\pi] \qquad (9)$$

where the unknown $\overline{\epsilon}$ characterizes the strain from a circle with outer radius R to an ellipse with longer and shorter semi-axes $R(1 + \overline{\epsilon})$ and $R(1 - \overline{\epsilon})$, and $\mathbf{i}$ and $\mathbf{j}$ are the unit vectors along the major and minor axes, respectively.

Similar to the case of plane strain bulk, the two ends of a CNT were constrained axially. Then following similar steps to the case of plane strain bulk's modulus, the in-plane shear modulus can be determined using the following relation [5]:

$$U = 2G_{23}\overline{\epsilon}'^2 \qquad (10)$$

From the above five independent elastic moduli, any desired elastic coefficient can be obtained [62]; such that

$$E_2 = E_3 = \frac{4G_{23}K_{23}}{K_{23} + \psi G_{23}} \qquad (11)$$

$$\nu_{23} = \frac{K_{23} - \psi G_{23}}{K_{23} + \psi G_{23}} \qquad (12)$$

$$\psi = 1 + \frac{4K_{23}\nu_{21}^2}{E_1} \qquad (13)$$

where $E_2$ and $\nu_{23}$ refer to the transverse Young's modulus and Poisson's ratio, respectively.

## 3 Results and discussion

In order to verify the validity of current MD simulations, the five independent elastic coefficients of pristine CNTs determined by Shen and Li [2] are considered. They derived closed-form solutions for all five independent elastic coefficients using an energy approach in the framework of molecular mechanics. Table 2 summarises the outcome of this comparison. The two sets of results predicted by different modeling techniques are in good agreement, validating the MD simulations carried out in this study. The differences observed for larger diameter CNTs, especially for the values of $K_{23}$ and $G_{23}$ are attributed to the fact that Shen and Li [2] did not consider the force constants corresponding to bond torsion and vdW interactions, whereas AIREBO force field used herein accounts these terms.

From the way of its folding, the nanotube can be identified as armchair or zigzag nanotube. We considered both types of CNTs having almost identical diameters to study the effect of chirality on the mechanical behavior. Figure 3 demonstrates the variation of PE of armchair and zigzag CNTs subjected to axial strain. As expected the larger diameter CNTs show higher PE energy than smaller ones. It may also be observed that both types of CNTs show almost similar trends till they were fractured. The marginal differences are attributed to the variation in their diameters and chirality. For the sake brevity, we considered only armchair CNTs to study the effect of vacancies on their elastic properties. Subsequently, we performed MD simulations on pristine and vacancy defected armchair CNTs until they fractured under the axial loading, as shown in Fig. 4. Vacancy defects were introduced in such a way that the percentage of defects is the same in the CNTs irrespective of their geometrical configurations. The same MD modeling procedure detailed in the preceding section is adopted to determine the elastic coefficients of defective CNTs. It may be observed from Fig. 4 that the defected CNTs were fractured at low strain levels than pristine CNTs and larger vacancy concentration ($> 2\%$) may significantly affects the mechanical behaviour of CNTs. The sets of energy-strain and stress-strain curves obtained in the current study are found to be in good agreement with those obtained for pristine CNTs in the existing MD studies [17, 22, 63].

We performed comprehensive MD simulations to determine the five independent elastic coefficients of pristine and vacancy defected CNTs subjected to different types of loadings. First, the pristine CNTs were considered and their elastic properties were determined, as shown in Fig. -9. This is evident from results that the elastic coefficients of the CNTs decrease as their diameters increase. This attributes to the variation of included angles of two adjacent C−C bonds of a CNT with its radius. When a graphene sheet forms nanotube structure, the lengths and spatial relations of C−C bonds change obviously. Therefore, the influence of included angles of CNTs decreases as their diameter increase and the elastic properties of larger CNTs decrease and become closer to that of graphene. Following that, vacancies were introduced in such a way that their concentration is the same in the CNTs irrespective of their geometrical configurations. Table 1 summarizes the total number of atoms in the CNTs prior to and post the introduction of vacancies. Fig. demonstrates the effect of vacancies on the axial Young's modulus ($E_1$) of CNTs. As the vacancy concentration increases, the values of $E_1$ of the smaller diameter defective CNTs decrease more rapidly in comparison with the larger ones. The decrease in the value of $E_1$ of (25, 25) defective CNT remains constant; (4%) irrespective of the vacancy concentration. This is again attributed to the fact that the influence of the included angles of two adjacent C−C bonds in CNTs decreases as their diameter increase, and hence the fluctuation of energy differences for larger diameter defective CNTs are found to be marginal for the same vacancy concentration and hence show no degradation. The findings for smaller defective CNTs are consistent with the previously reported results of Refs. [13, 16, 20]. For instance, the decrease in Young's moduli of (5, 5) defective CNTs with 1% and 2% vacancy concentrations are reported by Yuan and Liew [13] as 7.9% and 15.8%, respectively. Likewise, a corresponding decrease in Young's moduli was reported by Rafiee and Pourazizi [20] of 7.1% and 12.5%, respectively. For (10, 10) CNT with 1% vacancy concentration, the percentage reductions in Young's moduli are reported as 8.7% and 6% in Refs. [16] and [20], respectively. In the current study, the percentage reduction is 7.14%. The influence of vacancies on the major Poisson's ($v_{12}$) ratios is marginal irrespective of vacancy concentration, as illustrated in Fig. . A similar trend of results is obtained for the values of $G_{12}$ (see Fig. ) but in this case the effect of vacancies is seen to be more pronounced. This finding is also consistent with the results reported by Yang et al. [26] and Sharma et al. [27]. In their study, the influence of defects is more dominant on the shear moduli compared to Young's moduli. Fig. illustrates the effect of vacancies on the plane strain bulk's ($K_{23}$) modulus of CNTs. The vacancies significantly affect the values of $K_{23}$, and their influence become more prominent for larger diameter CNTs, in contrast to the cases of $E_1$ and $G_{12}$. Similar trend observed for the in-plane shear ($G_{23}$) modulus of CNTs as demonstrated in Fig. , but the influence of vacancies is less pronounced.

In the previous sets of results, the elastic properties of defective CNTs were determined considering vacancies in their middle portions. However, the variation of the position of vacancies along the length of CNT may influence the elastic properties of CNTs. Therefore, the position of vacancies was moved along the axis of a CNT in nine steps: $\pm x/L = 0.1$, $\pm x/L = 0.23$, $\pm x/L = 0.3$, $\pm x/L = 0.4$, and $x/L = 0$. Fig. *2* demonstrates schematically the position of vacancies along the molecular structures of (5, 5) CNTs. **Error! Reference source not found.** shows the influence of the position of vacancies on the values of $E_1$, $G_{12}$ and $K_{23}$. These elastic moduli are normalized with the moduli of pristine CNTs for both vacancy concentrations. We can observe that the values of $E_1$ and $G_{12}$ increases as the position of vacancies move from the end to the centre of CNTs, especially for smaller diameter of CNTs. On the other hand, it is evident from **Error! Reference source not found.**1 (e) and (f) that the position of vacancies does not affect the values of $K_{23}$. Although not shown here the same trend is observed in case of $G_{23}$. This is due to the fact that, irrespective of the position of vacancies, the transverse deformation was found to be almost the same while determining the averaged $K_{23}$ and $G_{23}$ elastic coefficients.

## 4 Conclusions

As the first of its kind, this study reports the transversely isotropic elastic properties of defective CNTs and investigates the effects of position of vacancies on their elastic properties within the framework of MD simulations. Results were compared with pristine CNTs for analyzing the influence mechanisms of vacancies and their positions. Good agreement between the current results and the existing results obtained via different approaches for the pristine CNTs and the axial elastic properties ($E_1$ and $G_{12}$) of defective CNTs validate the accuracy of employed MD modeling. In this paper, the influence of vacancies on the transversely isotropic elastic of CNTs is studied for the first time. The following is a summary of the significant findings:

1. The plane strain bulk and in-plane shear moduli of defective CNTs decrease more rapidly in comparison to all other elastic coefficients for the same vacancy concentration. On the other hand, vacancies marginally influence the axial Young's modulus of CNTs.

2. The effect of vacancies on the axial Young's and shear moduli is found to diminish as a CNT diameter increases and the converse is true for the plane strain bulk and in-plane shear moduli.

3. The position of vacancies along the length of CNTs significantly affects the axial elastic properties, especially those of smaller diameter CNTs.

**Table 1** Geometrical parameters of armchair CNTs

| Chirality | Length (Å) | Diameter (Å) | Number of atoms | | |
|---|---|---|---|---|---|
| | | | $\rho =$ | $\rho = 1\%$ | $\rho = 2\%$ |
| (5, 5) | 33.90 | 6.781 | 280 | 277 | 274 |
| (10, 10) | 67.81 | 13.563 | 1120 | 1109 | 1098 |
| (15, 15) | 101.22 | 20.344 | 2520 | 2495 | 2470 |
| (20, 20) | 135.62 | 27.125 | 4480 | 4435 | 4390 |
| (25, 25) | 339.1 | 33.339 | 13900 | 13761 | 13622 |

**Table 2** Comparison of the five independent engineering constants of pristine CNTs

| Chirality | Diameter (Å) | Ref. | $E_1$ (TPa) | $\upsilon_{12}$ | $G_{12}$ (TPa) | $K_{23}$ (TPa) | $G_{23}$ (TPa) |
|-----------|--------------|---------|-------|-------|-------|-------|-------|
| (5, 5)    | 6.781        | Present | 2.12  | 0.174 | 0.802 | 0.54  | 0.14  |
|           |              | [2]     | 2.08  | 0.172 | 0.791 | 0.536 | 0.132 |
| (10, 10)  | 13.563       | Present | 1.2   | 0.17  | 0.445 | 0.391 | 0.125 |
|           |              | [2]     | 1.06  | 0.162 | 0.442 | 0.271 | 0.017 |
| (15, 15)  | 20.344       | Present | 0.952 | 0.165 | 0.31  | 0.265 | 0.105 |
|           |              | [2]     | 0.707 | 0.161 | 0.301 | 0.181 | 0.005 |
| (20, 20)  | 27.125       | Present | 0.905 | 0.163 | 0.24  | 0.21  | 0.08  |
|           |              | [2]     | 0.531 | 0.16  | 0.227 | 0.136 | 0.002 |

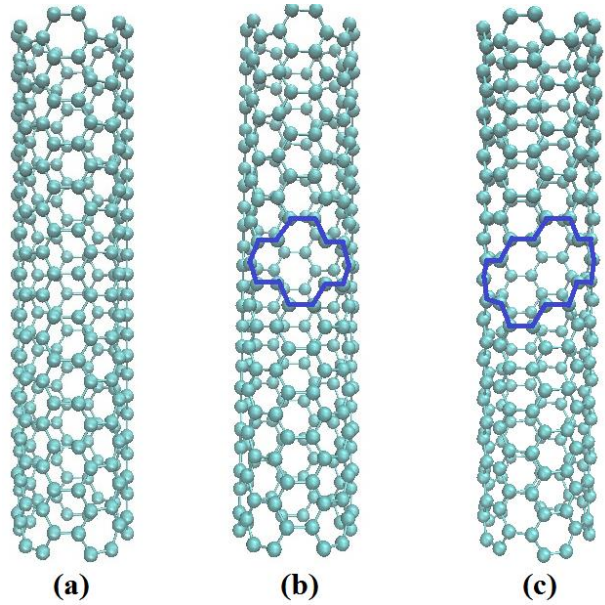

**Fig. 1** Schematics of different vacancies adopted in the study: (a) pristine (5, 5) CNT, (b) defective (5, 5) CNT ($\rho = 1\%$), and (c) defective (5, 5) CNT ($\rho = 2\%$)

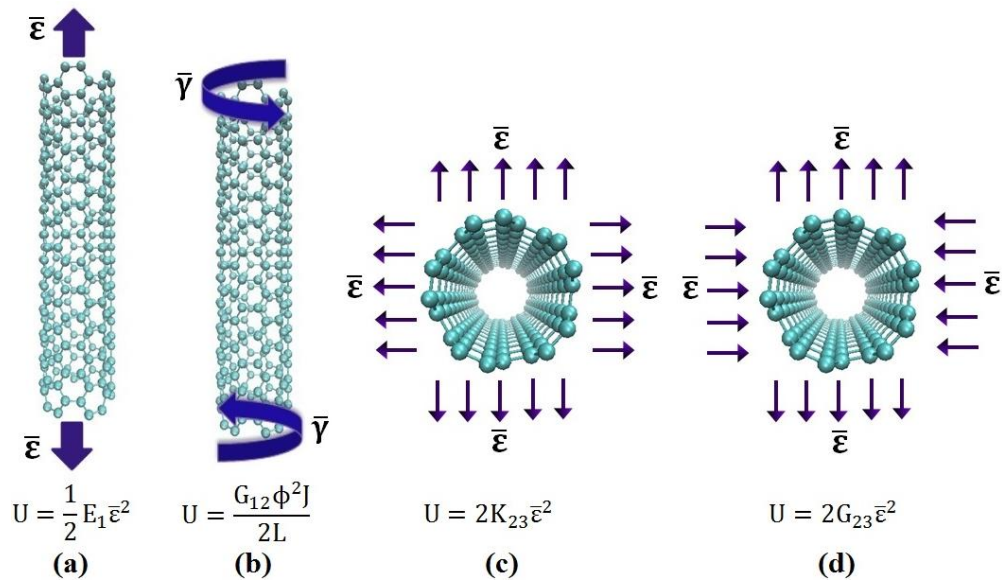

$$U = \frac{1}{2}E_1\bar{\varepsilon}^2 \qquad U = \frac{G_{12}\phi^2 J}{2L} \qquad U = 2K_{23}\bar{\varepsilon}^2 \qquad U = 2G_{23}\bar{\varepsilon}^2$$

**(a)**      **(b)**      **(c)**      **(d)**

**Fig. 2** The four types of loading conditions: (a) tensile, (b) twist, (c) in-plane bi-axial tension, and (d) in-plane shear

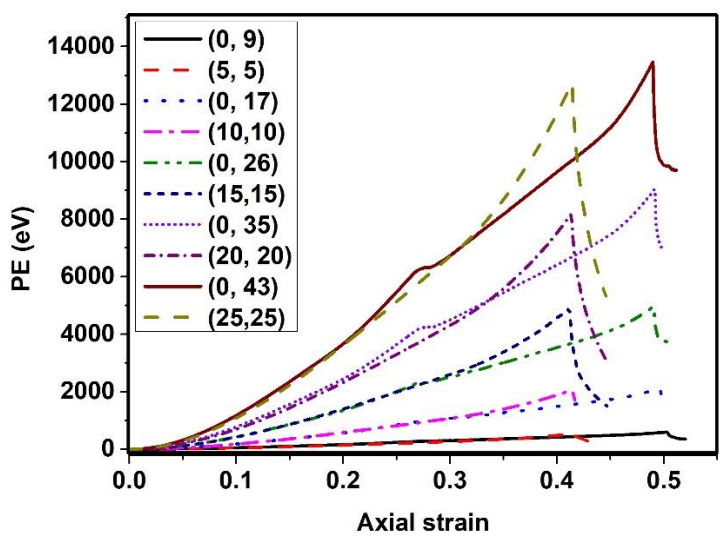

**Fig. 3** The variation of the PE of armchair and zigzag CNTs

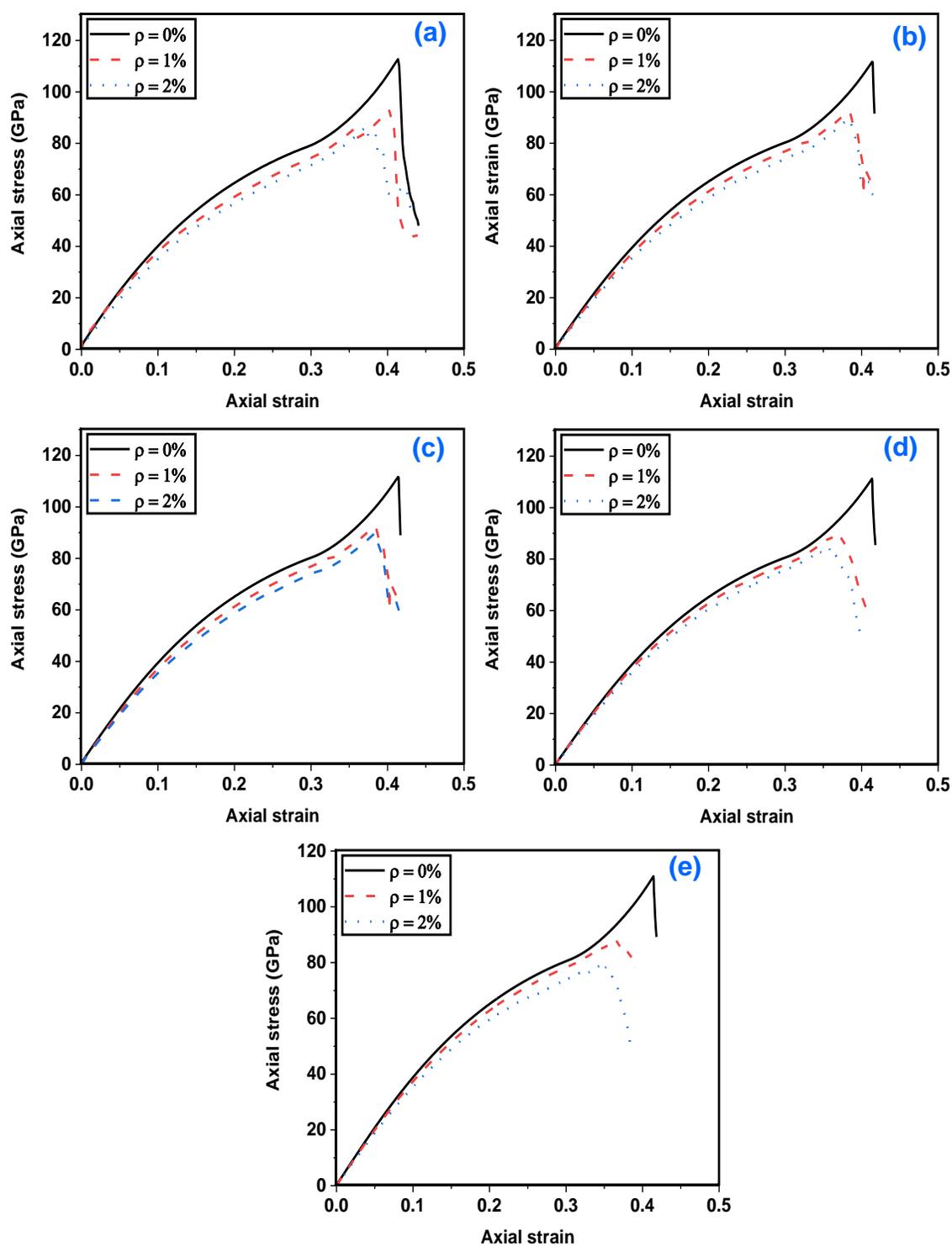

**Fig. 4** Stress-strain curves for pristine and vacancy defected armchair CNTs: (a) (5, 5), (b) (10, 10), (c) (15, 15), (d) (20, 20), and (e) (25, 25)

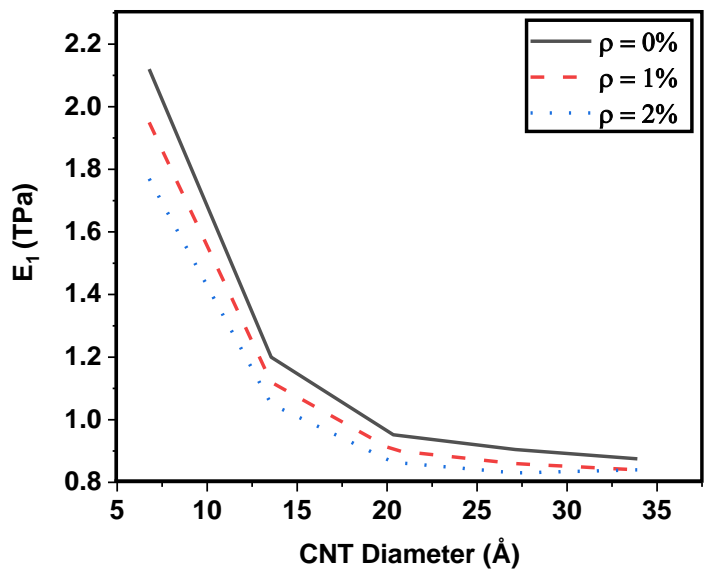

**Fig. 5** Effect of vacancy defects on the axial Young's moduli of different armchair CNTs

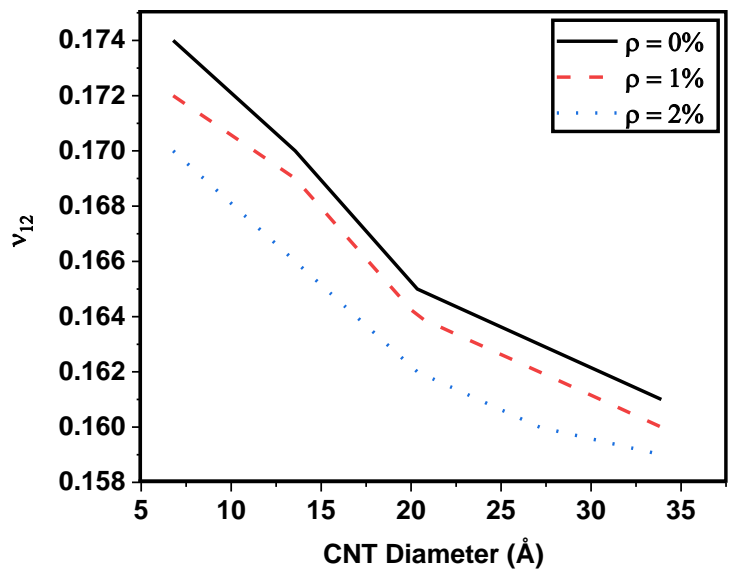

**Fig. 6** Effect of vacancy defects on the major Poisson's ratios of different armchair CNTs

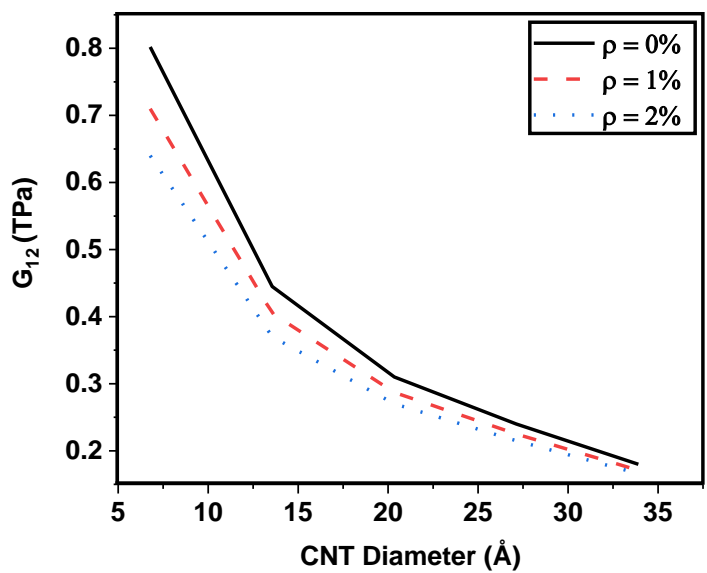

**Fig. 7** Effect of vacancy defects on the axial shear moduli of different armchair CNTs

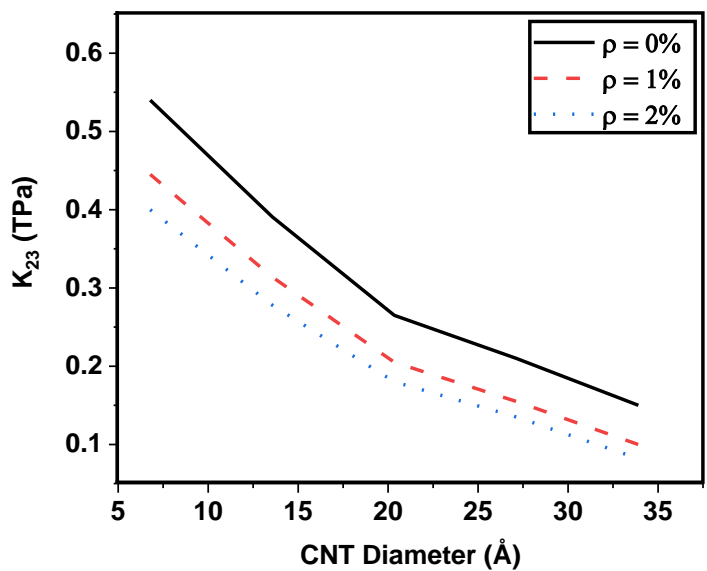

**Fig. 8** Effect of vacancy defects on the plane strain bulk moduli of different armchair CNTs

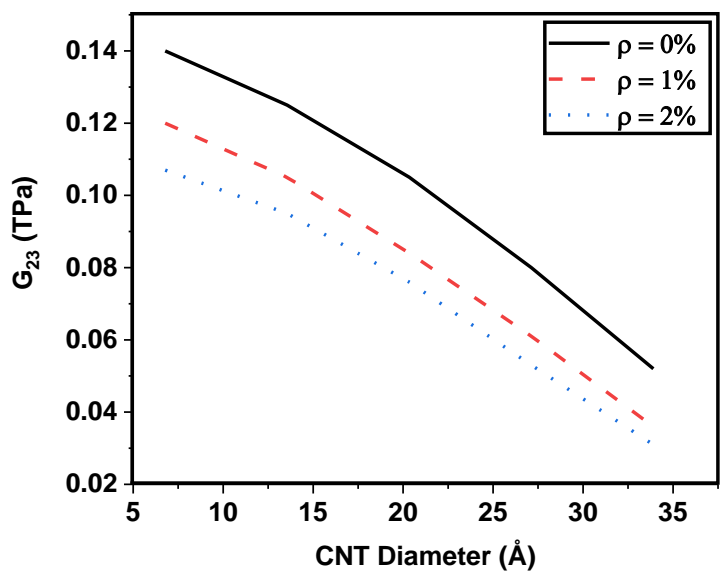

**Fig. 9** Effect of vacancy defects on the in-plane shear moduli of different armchair CNTs

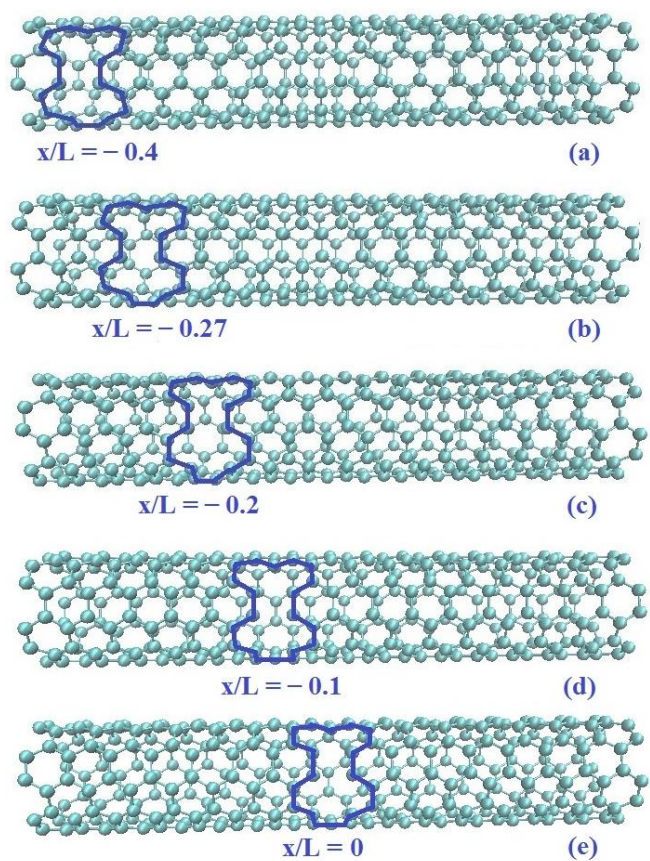

**Fig. 2** Position of vacancies along the length of (5, 5) CNT

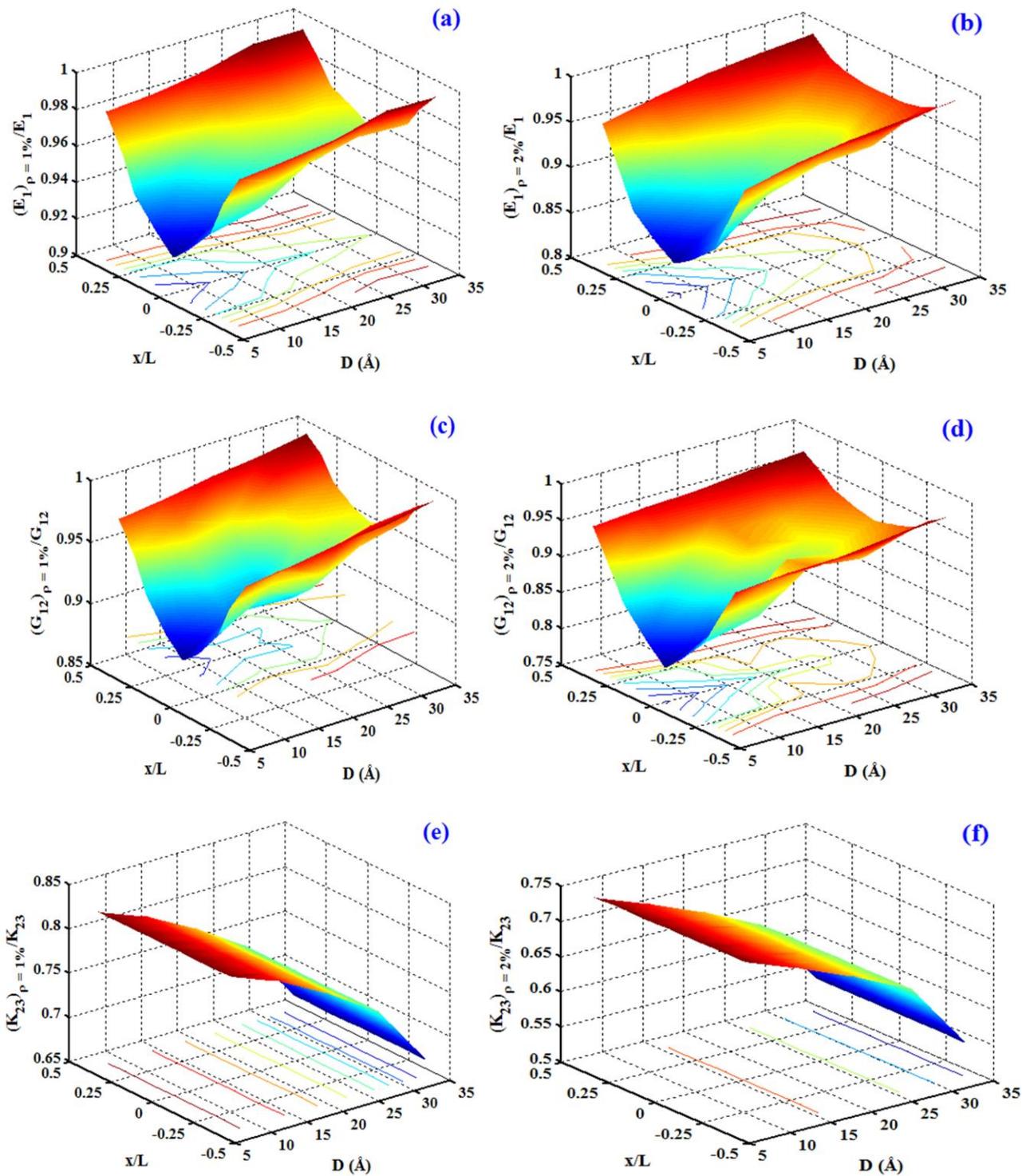

**Fig. 11** Effect of position of vacancies along the length of CNTs on the normalized elastic coefficients: (a) $E_1$ ($\rho = 1\%$), (b) $E_1$ ($\rho = 2\%$), (c) $G_{12}$ ($\rho = 1\%$), (d) $G_{12}$ ($\rho = 2\%$), (e) $K_{23}$ ($\rho = 1\%$), and (f) $K_{23}$ ($\rho = 2\%$)